\documentclass{PoS}

\usepackage{amsmath}
\usepackage{amssymb}
\usepackage{dsfont}

\title{Finite Field Theories and Causality}

\ShortTitle{Causal Perturbation Theory}

\author{\speaker{Andreas Aste}\\

        Department of Physics, University of Basel\\
        Klingelbergstrasse 82, CH-4056 Basel, Switzerland\\
        E-mail: \email{andreas.aste@unibas.ch}}

\abstract{\vskip 0.2cm A condensed introduction to the basic concepts of causal
perturbation theory is given. Causal perturbation theory is a mathematically rigorous
approach to renormalization theory, which makes it possible to put the theoretical
setup of perturbative quantum field theory on a sound mathematical basis by avoiding
infinities from the outset.
It goes back to a seminal work by Henri Epstein and Vladimir Jurko Glaser published
in 1973, where a specific causality condition was imposed at every order of perturbation
theory in the case of scalar quantum field theory such that divergent integrals could be
avoided in actual calculations of loop diagrams.
In the meantime, the causal approach has been applied also to a wide range of
gauge theories. 
}

\FullConference{LIGHT CONE 2008 Relativistic Nuclear and Particle Physics\\
		 July 7-11  2008\\
		 Mulhouse, France}

\begin{document}

\section{Introduction}
Before we come to the actual problem of ultraviolet (UV) divergences
in perturbative quantum field theory (pQFT), we discuss here a
naive example of 'UV divergence'
by considering Heaviside-$\Theta$- and Dirac-$\delta$-distributions in 1-dim.
'configuration space'.
The product of these two distributions $\Theta(x) \delta(x)$ is
obviously ill-defined, however, considering the Fourier transforms
\begin{equation}
\mathcal{F} \{ \delta \} (k)=
\hat{\delta} (k) = \int dx \, \delta(x) e^{-ikx} =1,
\end{equation}
\begin{equation}
\hat{\Theta} (k) = \lim_{\epsilon \searrow 0}
\int dx \, \Theta(x) e^{-ikx-\epsilon x} = \lim_{\epsilon \searrow 0}
\frac{ie^{-ikx-\epsilon x}}{k-i \epsilon} \Biggr|^{\infty}_{0}= -\frac{i}{k-i0},
\end{equation}
one can nevertheless calculate in a formal manner the Fourier transform of the
ill-defined product mentioned above
\begin{equation}
\mathcal{F} \{ \Theta \delta \} (k)=\int dx \, e^{-ikx} \Theta(x)  \delta(x) =
\int dx \, e^{-ikx} \int \frac{dk'}{2 \pi} \hat{\Theta}(k') e^{+ik'x}
\int \frac{dk''}{2 \pi} \hat{\delta}(k'') e^{+ik''x}.
\end{equation}
Since $\int dx \, e^{i(k'+k''-k)x}=2 \pi \delta(k'+k''-k)$, we obtain the divergent
convolution integral
\begin{equation}
\mathcal{F} \{ \Theta \delta \} (k) = \frac{1}{2 \pi} \int dk' \,
\hat{\Theta} (k') \hat{\delta} (k-k') = {
-\frac{i}{2 \pi} \int \frac{dk'}{k'-i0}}.
\end{equation}
The obvious problem in x-space leads to a divergent integral in k-space.
This situation arises in a completely analogous manner in pQFT, as will be discussed
in the following.

\section{The origin of UV divergences in perturbative quantum field theory}
In pQFT, the r\^{o}le of the {Heaviside $\Theta$-distribution} is taken over
by the time-ordering operator. The well-known textbook expression for the 
perturbative scattering matrix given by
\begin{displaymath}
S  = \sum \limits_{n=0}^{\infty} \frac{(-i)^n}{n!}
\int \limits_{-\infty}^{+\infty} dt_1 \ldots 
\int \limits_{-\infty}^{+\infty} dt_n \, {T}
[H_{int}(t_1) \ldots H_{int}(t_n)]
\end{displaymath}
\begin{equation}
 =  \sum \limits_{n=0}^{\infty} \frac{(-i)^n}{n!}
\int  d^4 x_1 \ldots \int  d^4 x_n \, {T}
[\mathcal{H}_{int}(x_1) \ldots
\mathcal{H}_{int}(x_n)], \label{Smatrix_textbook}
\end{equation}
where the interaction Hamiltonian {$H_{int}(t)$}
is given by the interaction Hamiltonian density $\mathcal{H}_{int}(x)$ via
$H_{int}(t)=\int d^3 x \, 
\mathcal{H}_{int}(x)$, is problematic in the UV regime (and in the infrared
regime, when massless fields are involved).
A time-ordered expression \`a la
\begin{equation}
T [\mathcal{H}_{int}(x_1) \ldots \mathcal{H}_{int}(x_n)] = \! \!
\sum \limits_{Perm. \, \, \Pi} \Theta(x^0_{\Pi_1}-x^0_{\Pi_2}) \ldots
\Theta(x^0_{\Pi_{(n-1)}}-x^0_{\Pi_n}) \mathcal{H}_{int}(x_{\Pi_1})
\ldots \mathcal{H}_{int}(x_{\Pi_n})
\end{equation}
is formal (i.e., ill-defined), since the operator-valued distribution
products of the $\mathcal{H}_{int}$ are simply too singular to be multiplied by
$\Theta$-distributions. This observation provides a formal explanation
for the existence of UV divergent expressions in pQFT, which are usually
related in a qualitative manner to contributions of virtual particles with
'very high energy', or, equivalently, to physical phenomena at very short distances.

A first step towards the solution of the apparent mathematical problem
was taken by N.~N.~Bogoliubov and D.~V.~Shirkov \cite{Bogol}, which
introduced a clear definition of the causality condition in pQFT
and the concept of adiabatic switching (see below). However, in their attempt
the UV divergences did persist. A rigorous mathematical analysis within the
framework of distribution theory was finally presented by
H.~Epstein and V.~Glaser \cite{Epstein}, who derived an inductive
construction of the perturbation series based on Poincar\'e invariance and causality
(unitarity plays no immediate r\^ole). In their approach,
UV divergences are avoided from the start, and the Feynman rules only hold on tree-level.
The calculation of loop diagrams turns out to be rather technical and involves
finite (subtracted) dispersion integrals instead of divergent Feynman integrals.
A new strategy to treat the infrared problem by adiabatic switching of the interaction
is also introduced in the causal approach.

In the following, an introduction to the basic concepts of the causal Epstein-Glaser
method is given. It is important to note that the presentation below suffers from
a certain lack of mathematical rigor as a natural consequence of the limited
space available in these proceedings.
For a concise introduction to the causal method we refer to the textbook
of G.~Scharf \cite{Scharf}.

\section{Mathematical preliminaries}
\subsection{Free quantum fields and operator valued distributions}
It is a crucial observation that free field operators are operator-valued
\emph{distributions}. E.g., for a scalar (neutral) field of a particle
with mass $m$ one has
\begin{equation}
\varphi(x)=\frac{1}{(2 \pi)^{3/2}} \int \frac{d^3 k}{\sqrt{2 E}}
\Bigl[ a(\vec{k}) e^{-ikx} + a^+ (\vec{k}) e^{+ikx} \Bigr], \quad E=\sqrt{
\vec{k}^2+m^2},
\end{equation}
where the annihilation- and creation operators
$a$ and $a^+$ fulfill the usual commutation relations.
This expression must be smeared out by rapidly decreasing test functions
$g(x)$ in the Schwartz space $\mathcal{S}(\mathds{R}^4)$
($\mathcal{S}$ ($\mathds{R}^3)$ would also be sufficient in the case of a free field)
in order to get an operator in Fock space, formally written in integral form as
$\varphi(g)=\int d^4x \, \varphi (x) g(x)$. 
Note that $\varphi(x) |0\rangle$ is not a Fock state.
The same arguing applies to the interaction Hamiltonian densities
used in perturbation theory constructed from normally ordered products of
free fields. E.g., in QED the perturbative interaction Hamiltonian
density is expressed by the help of the free spinor field $\Psi (x)$ and
the photon field $A_\mu (x)$ by the well-defined operator-valued
distribution $\mathcal{H}_{int}=-e : \bar{\Psi}(x) \gamma^\mu \Psi(x) : A_\mu(x)$.
In $\varphi^3$-theory, which will serve as a model theory in the following,
one has $\mathcal{H}_{int}=\frac{\lambda}{3!} :\varphi(x)^3:$.

\subsection{The perturbative S-matrix: basic properties}
Based on the observations made above, it is therefore most natural to replace the
problematic expression eq. (\ref{Smatrix_textbook}) by
\begin{equation}
S(g) =\mathds{1}+\sum \limits_{n=1}^{\infty} \frac{1}{n!}
\int  d^4 x_1 \ldots  d^4 x_n \,
{T_n (x_1, \ldots, x_n) g(x_1) \ldots g(x_n)}, \quad
{g \! \in \! \mathcal{S}(\mathds{R}^4)}, \label{pertS}
\end{equation}
where the switching function $g(x)$ can be considered as a local variation
of the coupling constant and we set $T_1(x)=-i \mathcal{H}_{int}(x)$.
In the causal approach, the $T_n(x_1,\ldots,x_n) \simeq T [T_1(x_1) \ldots T_1(x_n)]$
denote a well-defined (divergence free) time-ordered product, which is symmetric in the
sense that
\begin{equation}
T_n(\ldots,x_{{i}},\ldots,x_{{j}},\ldots)=
T_n(\ldots,x_{{j}},\ldots,x_{{i}},\ldots)
\quad \forall i,j
\end{equation}
by construction.
Since the $T_n$ are constructed such that they are free of any UV divergences,
every order of the perturbative $S-$matrix $S_n(g) = \frac{1}{n!}
\int  d^4 x_1 \ldots  d^4 x_n \,
{T_n (x_1, \ldots, x_n) g(x_1) \ldots g(x_n)}$
is well-defined even when massless fields are present. Infrared divergences
are absent as long as the interaction is switched by $g \in \mathcal{S}(\mathds{R}^4)$.
Infrared problems arise in the so-called adiabatic limit
{$g(x) \rightarrow 1$}, since $1 \not \in  \mathcal{S}(\mathds{R}^4)$.
Performing the adiabatic limit is a delicate task as far as
existence and uniqueness of the limit are concerned. The
limit has to be taken such that observable quantities (like cross sections)
remain finite. A typical strategy is to rescale $g(x)$
according to
\begin{equation}
\lim_{\epsilon \searrow 0} g(\epsilon x) \, \, \, \, "\!
\rightarrow" \, \, \, \, g(x)=g(0)=const. 
\end{equation}
No further regularizations, like the introduction of a finite photon in mass in
theories like $QED$, are then necessary in this approach.

One should note that no statements about the convergence of the full series
eq. (\ref{pertS}) can be made in general.
We further mention that perturbative expansion of the inverse perturbative $S$-matrix
is given by
\begin{equation}
S(g)^{-1}  =  \mathds{1}+\sum \limits_{n=1}^{\infty} \frac{1}{n!}
\int  d^4 x_1 \ldots  d^4 x_n \,
{\tilde{T}}_n (x_1, \ldots, x_n) g(x_1) \ldots g(x_n) = 
{(\mathds{1}+T)^{-1}=\mathds{1}+\sum \limits_{r=1}^{\infty} (-T)^r},
\end{equation}
where ${\tilde{T}}_n(X)=\sum \limits_{r=1}^{n} (-1)^r \sum \limits_{P_r}
T_{n_1}(X_1) \ldots T_{n_r} (X_r)$,
and {$X=\{ x_1, \ldots , x_n \}$}
is a disordered set and {$\sum \limits_{P_r}$
denotes {all partitions of $X$} into $r$ disjoint subsets
\begin{equation}
X=X_1 \cup \ldots \cup X_r, \quad X_j \ne  \emptyset ,
\quad X_i \cap X_j = \emptyset ,  \quad |X_j|=n_j.
\end{equation}

\section{The method of Epstein and Glaser}
\subsection{Inductive construction of the perturbative S-matrix}
Causality is the pivotal point in the UV divergence-free approach of Epstein
and Glaser. Causality is expressed by the condition
that if the switching function $g(x)=g_1(x)+g_2(x)$ can be decomposed such that
the supports of $g_1(x)$} and $g_2(x)$ (denoted by $\mbox{supp}(g_1)$ and
$\mbox{supp}(g_2)$) are space-like separated, i.e. if there
exists a reference frame such that from
$x \! \in \! \mbox{supp}(g_1)$ follows  $x^0<0$ and from
$y \! \in \! \mbox{supp}(g_2)$ follows $y^0>0$ (see Fig. 1),
then one has
$S(g_1+g_2)=S(g_2)S(g_1)$ for all switching functions
$g_1,g_2$ which fulfill the condition above (denoted by
$\mbox{supp}(g_1) < \mbox{supp}(g_2)$, where the symbol '$<$' should be
read as 'earlier').\\
This implies
$T_n(x_1,\ldots,x_n)=T_m(x_1,\ldots,x_m)T_{n-m}(x_{m+1},\ldots,x_n)$
$\, \, $if {$\{ x_1,\ldots,x_m \} > \{x_{m+1},\ldots,x_n\}$},
a condition  which is, of course, intuitively clear.

Note that the support of a (test) function $\mbox{supp}(g)$ is the closed set
obtained by taking the complement of the largest open set on which $g$ vanishes.
Thus we adopt an analogous definition of the support for distributions, which is
the complement of the largest open set on which a distribution vanishes.
A distribution vanishes on an open set if it vanishes for all test functions
whose supports are in the open set.
\begin{center}
\includegraphics[scale=0.24]{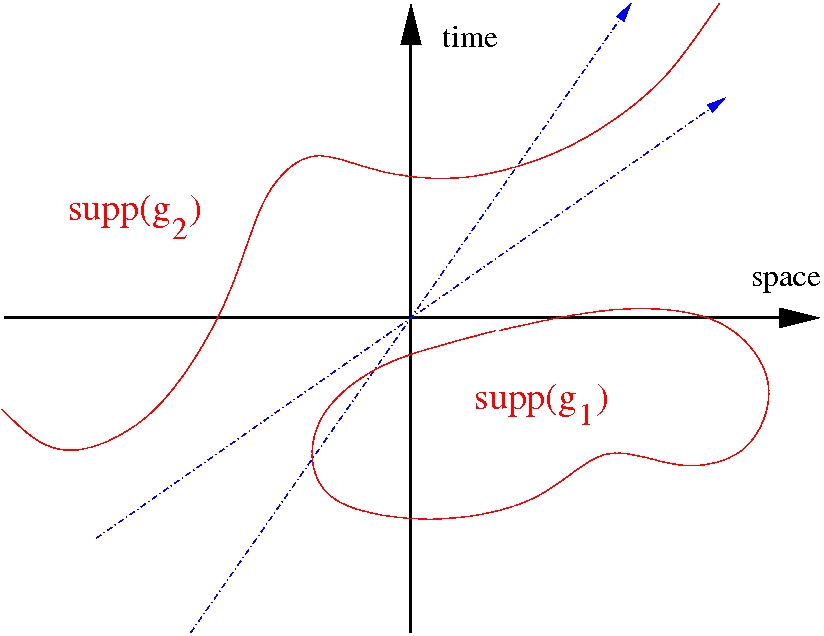}\\
Fig. 1: An example for $\mbox{supp}(g_2) > \mbox{supp}(g_1)$.
\end{center}
\subsection{Explicit construction of $S_2(g)$}
Causality and translation invariance require that
the commutator {$D_2(z=x_1-x_2)$}
\begin{equation}
{D_2(x_1-x_2)}=
(-i)^2 [\mathcal{H}_{int}(x_1),\mathcal{H}_{int}(x_2)]=
{
[T_1(x_1),T_1(x_2)]=0} \quad \mbox{for} \quad {(x_1-x_2)^2<0}
\end{equation}
has causal support on the (closed) four-dimensional light-cones
$\mbox{supp}(D_2)=\bar{V}^+(0) \cup \bar{V}^-(0)$ (see Fig. 2),
which are defined by
\begin{equation}
\bar{V}^+(x)=\{ y \, | \,  (y-x)^2 \ge 0, \, y^0 \ge x^0 \}, \quad
\bar{V}^-(x)=\{ y \, | \, (y-x)^2 \ge 0, \, y^0 \le x^0 \}. \label{lightcone4}
\end{equation}
One therefore introduces (primed) advanced and retarded distributions
$A_2^{{(')}}(z)$ and $R_2^{{(')}}(z)$ according to
\begin{eqnarray}
{R_2  =  +D_2 \Big|_{\bar{V}^+ - \{ 0 \}}}  & , & \,
{A_2  =  -D_2 \Big|_{\bar{V}^- - \{ 0 \}}}  , \quad
R_2, \, A_2=0 \, \, \, \mbox{elsewhere,}\\
{R'_2  =  -T_1(x_2)T_1(x_1)} & , & \,
{A'_2  =  -T_1(x_1)T_1(x_2)}  .
\end{eqnarray}
The non-trivial (!) splitting of $D_2$ into the retarded and advanced distributions
$R_2$ and $A_2$ (see Fig. 2) indeed corresponds to time-ordering: One has
$T[T_1(x_1) T_1(x_2)] \, "\!=\!" \, T_2(x_1,x_2)=R_2-R'_2=A_2-A'_2$.
One may substantiate this observation by explicitly checking that
$R_2-R'_2$ is given for
$z^0>0$ by $T_1(x_1)T_1(x_2)-T_1(x_2)T_1(x_1) +T_1(x_2)T_1(x_1)$, and for
$z^0<0$ by $+T_1(x_2)T_1(x_1)$.
However, the decomposition is not unique in general in the critical point
$z=0$.
\begin{center}
\includegraphics[scale=0.24]{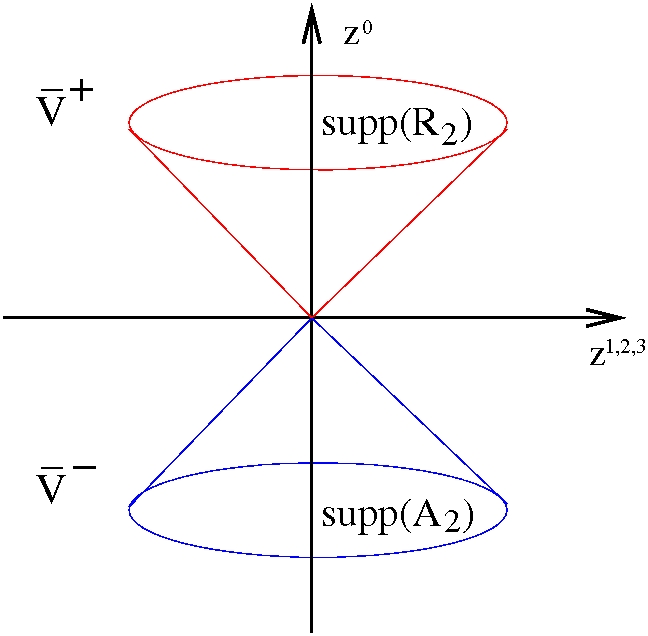}\\
Fig. 2: $D_2(z)=R_2(z)-A_2(z)=R'_2(z)-A'_2(z)$
\end{center}
Considering $\varphi^3$-theory as an example
with {$T_1(x)= \frac{i \lambda}{3!} :\varphi(x)^3:$}
and the scalar Feynman propagator defined via
$\langle0| T( \varphi(x_1) \varphi(x_2) ) |0\rangle =i \Delta_F(x_1-x_2)$,
standard Wick ordering leads to
\begin{displaymath}
T_2(x_1,x_2)  \,  "\!=\!" \, 
-\frac{\lambda^2}{3!^2} :\varphi(x_1)^3 \varphi(x_2)^3:
-\frac{9 \lambda^2}{3!^2} :\varphi(x_1)^2 \varphi(x_2)^2:
{i \Delta_F(x_1-x_2)}
\end{displaymath}
\begin{equation}
-\frac{18 \lambda^2}{3!^2} :\varphi(x_1) \varphi(x_2):
{[i \Delta_F(x_1-x_2)]^2}
-\frac{\lambda^2}{3!} {[i \Delta_F(x_1-x_2)]^3}.
\end{equation}
However, in the causal approach one constructs first
\begin{equation}
D_2(x_1-x_2)  =  -\frac{\lambda^2}{3!^2} [:\varphi(x_1)^3: , :\varphi(x_2)^3:]
 =  \ldots -\frac{9 \lambda^2}{3!^2} :\varphi(x_1)^2 \varphi(x_2)^2:
{i \Delta(x_1-x_2)}+ \ldots ,
\end{equation}
where $\Delta(x_1-x_2)$ is the {Pauli-Jordan distribution},
which can be decomposed into the {positive- and negative-frequency}
Pauli-Jordan distributions $\Delta(z)=\Delta^+(z)+\Delta^-(z)$ given by
\begin{equation}
\Delta^\pm (z)=\mp \frac{i}{(2 \pi)^3} \int d^4 k \, \Theta(\pm k^0) \delta(k^2-m^2)
e^{-ikx}.
\end{equation}

\subsubsection{Tree level}
In order to get the retarded C-number part of
$d_2^{tree}(z):=\Delta(z)$ of the particle-particle scattering diagram, we simply
multiply by $\Theta(z^0)$:
$r_2^{tree}(z)=\Theta(z^0) \Delta(z)$.
From $\hat{\Theta}(k)=\frac{i (2 \pi)^3}{k^0+i0}
\delta^{(3)}(\vec{k})$ follows (in a 'sloppy' style) in momentum space
\begin{equation}
\hat{r}_2^{tree}((k^0,\vec{0}))=\frac{i}{2 \pi}
\int \, dp^0 \frac{\hat{\Delta}((p^0,\vec{0}))}{k^0-p^0+i0}=
\frac{i}{2 \pi} \int dt \, \frac{\hat{\Delta}((tk^0,\vec{0}))}{1-t+i0},
\end{equation}
or, from the {Lorentz covariance} of $\Delta(z)$
($\hat{\Delta}(k)$)
one may derive a dispersion relation 
\begin{equation}
\hat{r}_2^{tree}(k)=\frac{i}{2 \pi} \int dt \,
\frac{\hat{\Delta}(tk)}{1-t+i0} \quad \mbox{for} \, \, k \! \in \! V^+.
\end{equation}
From
$\hat{\Delta}(k)=-2 \pi i \, \mbox{sgn} (k^0) \delta(k^2-m^2)$ follows
\begin{displaymath}
\hat{r}_2^{tree}(k)=\int dt \frac{\mbox{sgn}(tk^0) \delta(t^2 k^2-m^2)}{1-t+i0}
\end{displaymath}
\begin{equation}
=\int dt \frac{ [\delta(t-\frac{m}{\sqrt{k^2}})-\delta(t+\frac{m}{\sqrt{k^2}})] }
{2 \sqrt{k^2} m (1-t+i0)}={\frac{1}{k^2-m^2}} \quad \quad (k \! \in \! V^+).
\end{equation}
The full expressions for $\hat{r}_2^{tree}(k)$ and $\hat{t}_2^{tree}(k)$ for
arbitrary $k$ follow from analytic considerations.

\subsubsection{Loop level}
The self-energy (one-loop) part in $T_2$
is logarithmically divergent:
\begin{equation}
t_2^{loop}(x_1-x_2) \sim [i \Delta_F(x_1-x_2)]^2  \,
\longrightarrow^{\! \!  \! \! \!  \!\! \!  \! \! \!  \! \!  \!\mathcal{F}} \,
\int \frac{d^4 p}{[p^2-m^2+i0] [(k-p)^2-m^2+i0]}.
\end{equation}
In the causal approach,
one calculates first $D_2(x_1-x_2)=[T_1(x_1),T_1(x_2)]$, leading to
\begin{equation}
d_2^{loop}(x_1-x_2) \sim [\Delta^-(x_1-x_2)]^2-[\Delta^-(x_2-x_1)]^2 \,
\longrightarrow^{\mathcal{ \! \! \! \! \! \! \!  \! \! \!  \! \! \!  \!F}} \, \, \,
\mbox{sgn}(k^0) \Theta(k^2-4m^2).
\end{equation}
Naive splitting leads to a divergent dispersion integral, as in the
Feynman integral case:
\begin{equation}
\Theta(z^0) d_2^{loop}(z) \,
\longrightarrow^{ \! \! \! \! \! \! \!  \! \! \! \! \! \! \! \mathcal{F}} \, \,
\int dt \, \frac{\hat{d}_2^{loop}(tk)}{1-t+i0}
\quad (k \! \in \! V^+).
\end{equation}
However, it can be shown that the retarded part of $d_2^{loop}$ can be obtained
from a finite, subtracted dispersion integral (in the massive case $m \ne 0$)
\begin{equation}
\hat{r}_2^{loop}(k)=\frac{i}{2 \pi} \int \limits_{-\infty}^{+\infty}
\frac{\hat{d}_2^{loop}(tk)}{(t-i0)^{\omega+1}
(1-t+i0)}+{const.}
\quad (k \! \in \! V^+) \quad {\mbox{with}} \, \, 
\omega=0.
\end{equation}
$\omega$ derives from the scaling properties
($\rightarrow$ power counting degree of divergence)
of the causal distribution $d_2^{loop}$.

\subsection{Higher orders}
With causality as the fundamental input,
it is possible to construct causal distributions\\
$A_n, \, R_n \, \, \mbox{and} \, \, D_n(x_1,\ldots,x_n)=
R_n(x_1,\ldots,x_n)-A_n(x_1,\ldots,x_n)$
at every order of perturbation theory, which have causal support
\begin{equation}
\mbox{supp} \, R_n(x_1,\ldots,{x_n})  \subseteq   \Gamma^+(x_n),\quad
\mbox{supp} \, A_n(x_1,\ldots,{x_n})  \subseteq   \Gamma^-(x_n),
\end{equation}
where the generalized closed forward (backward) light-cones
are defined by
\begin{equation}
\Gamma^\pm(x_{n})=\{(x_1,\ldots,x_{n}) \, | \, x_j \! \in \! \bar{V}^\pm (x_n)
\, \forall \, j=1,\ldots,n-1 \}.
\end{equation}
Obviously, one has $\mbox{supp} \, D_n(x_1,\ldots,x_n)
\subseteq  \Gamma^+(x_n) \cup \Gamma^-(x_n)$.
We give here a short recipe how the higher order distributions are
constructed. The crucial step in the inductive construction of the
$T_n$ is the splitting of the $D_n$, which can be performed in
a mathematically well-defined ('finite') manner. However, the result
is in general not unique, a fact which is directly related to the possibility
of finite renormalizations of Green's functions in the standard
approaches to renormalization theory.
Recipe:
\begin{center}
\vskip -0.8cm $T_m$ known for $1 \le m \le n-1$
\begin{equation*}
%\vskip -1.3cm \Downarrow \vskip -1.8cm 
\Downarrow
\end{equation*}
construct primed distributions
\begin{equation*}
A'_n(x_1,\ldots,x_n)  =  \sum \limits_{P_2} \tilde{T}_{n_1}(X) T_{n-n_1}(Y,x_n)
\quad {and} \quad 
R'_n(x_1,\ldots,x_n)  =  \sum \limits_{P_2}  T_{n-n_1}(Y,x_n) \tilde{T}_{n_1}(X)
\end{equation*}
with $P_2:\{x_1,\ldots,x_{n-1} \}=X \cup Y, \, X \neq \emptyset , n_1=|X| \ge 1$
\begin{equation*}
\Downarrow
\end{equation*}
allow $X=\emptyset$, consider distributions with causal (light-cone) support
\begin{equation*}
A_n(x_1,\ldots,x_n)  =  A'_n(x_1,\ldots,x_n)+T_n(x_1,\ldots,x_n) \, , \quad 
R_n(x_1,\ldots,x_n)  =  R'_n(x_1,\ldots,x_n)+T_n(x_1,\ldots,x_n)
\end{equation*}
\begin{equation*}
\Downarrow
\end{equation*}
$T_n$ unknown, but difference distribution
\begin{equation*}
D_n=R'_n-A'_n=R_n-A_n
\end{equation*}
can be shown to be causal: distribution splitting of $D_n$ generates $R_n$
\begin{equation*}
\Downarrow
\end{equation*}
\begin{equation*}
T_n=R_n-R'_n \, !
\end{equation*}
\end{center}
As demonstrated above at second order for the particle-particle scattering
tree diagram and the self-energy loop diagram,
the C-number parts of the $D_n$, $R_n$:
$d^{tree,loop,\ldots}(x_1-x_n,\ldots,x_{n-1}-x_n)$,
$r^{tree,loop,\ldots}(x_1-x_n,\ldots,x_{n-1}-x_n)$
go over into
$\hat{r}^{tree,loop,\ldots}(p_1,\ldots,p_{n-1}) \, \, \mbox{and} \, \,
\hat{d}^{tree,loop,\ldots}(p_1,\ldots,p_{n-1})$
via Fourier transformation.
If at least one field in a field theory is
massive, it can be shown in general that at all orders of perturbation theory, for
$p=(p_1,p_2,\ldots) \! \in \! \Gamma^+$ a subtracted dispersion relation applies
as a distribution splitting operator
\begin{equation}
\hat{r}(p)=\frac{i}{2 \pi} \int \limits_{-\infty}^{+\infty}
\frac{\hat{d}(tp)}{(t-i0)^{\omega+1}(1-t+i0)} dt +
\sum \limits_{|\alpha|=0}^{\omega} c_\alpha p^\alpha,
\quad (\alpha: \mbox{multi-index}),
\end{equation}
where $\omega$ is a rigorously defined power-counting degree of divergence.
The terms $\sum c_\alpha p^\alpha$ correspond to possible finite renormalization
terms (which have been discussed in connection with the renormalization group
of QED in \cite{Scharf}).
The polynomial terms in Fourier space are due to the fact that the splitting of
$D_n$ into $R_n$ and $A_n$ is not uniquely defined at the "tip" $x_1=x_2=\ldots=x_n$
of the generalized forward/backward light-cones $\Gamma^\pm$.
In configuration space, they correspond to local terms
$\sim \sum \hat{c}_\alpha D^\alpha \delta(x_1-x_n,\ldots,x_{n-1}-x_n)$.
Perturbation theory alone does not specify them, and they have to be
restricted, e.g., by symmetry considerations and may lead to subsequent
finite renormalizations of the $T_n$, as already mentioned.

\section{Gauge theories}
We shortly mention the case of purely gluonic QCD as a standard example for a gauge theory.
Within the causal approach,
it is most natural to start from a first order gluon field
coupling (matter fields neglected)
\begin{equation}
T_1(x)=i \frac{g}{2} f_{abc} :A_\mu^a(x) A_\nu^b(x) F^{\nu \mu}_c(x):,
\quad F^{\mu \nu}_c (x)=\partial^\mu A^\nu_c (x)-\partial^\nu A^\mu_c (x).
\end{equation}
First order gauge invariance
requires additional fields (ghosts)
\begin{equation}
T_1(x)=i g f_{abc} :A_\mu^a(x) A_\nu^b(x) \partial^\nu A^\mu_c(x):
-ig f_{abc} : A^\mu_a(x) u_b(x) \partial^\mu \tilde{u}_c (x): \, .
\end{equation}
At second order, tree diagrams containing C-number
distributions
$\sim \partial_\mu \partial_\nu \Delta(x_1-x_2) {
\, \, \, \longrightarrow^{\! \! \! \! \! \! \! \!\! \! \! \! \! \!\mathcal{F}}} \, \, \,
-k^\mu k^\nu \hat{\Delta}(k)$
appear; their splitting is fixed up to a local term
$\sim g^{\mu \nu} \delta^{(4)}(x_1-x_2)$ according to distribution theory.
\begin{equation}
k^\mu k^\nu \hat{\Delta}(k) \, \, \, 
\longrightarrow^{ \! \! \! \! \! \! \! \! \! \! \! \! \! \! \! \! \! \! \! \!
\! \! \! \! \! splitting}
\, \, \, \frac{k^\mu k^\nu}{k^2+i0}+{C} g^{\mu \nu}.
\end{equation}
Second order gauge invariance determines the constant
$C$ such that the usual four-gluon coupling term is automatically generated \cite{Aste}.

\section{Concluding remarks}
We conclude by giving a short and incomplete review of recent activities
concerning the causal Epstein-Glaser approach. The causal approach was
'rediscovered' by Michael D\"utsch and G\"unter Scharf (U. Z\"urich) in 1985.
Their work resulted in a textbook \cite{Scharf}, where a complete discussion of
QED and introduction to the causal approach can be found.
From 1989-1993, the Z\"urich group led by G\"unter Scharf focused on, e.g.,
interacting fields in the causal approach \cite{Interacting},
axial anomalies \cite{Axial} and gave a full discussion of the renormalizability of
scalar QED \cite{Scalar}. During the period from 1993-1999, a complete discussion
of perturbative QCD was worked out \cite{Dutsch}, and gauge theories like the
full standard model (including the phenomenon of spontaneous symmetry breaking) were
studied \cite{Electroweak,Gracia}, theories in dimensions other than four
were considered \cite{Schwinger} and analytic calculations of multi-loop diagrams were
carried out \cite{Multiloop}.
Quantum gravity and supersymmetric theories were also considered,
and other groups (Klaus Fredenhagen et al.) generalized the causal approach to
field theories on curved space-times. Finally, recent work of Ernst
Werner and Pierre Grange in connection with light cone quantum field theory
should be mentioned, which will also be part of the LC2008 proceedings
(see also \cite{Werner}).

\end{document}